\documentclass[useAMS,usenatbib,usegraphicx]{mn2e}
\usepackage{epsf}

\title[Masses and thier evolution of high-$z$ galaxies in the
CDM model]{
Masses of high-$z$ galaxy hosting haloes from angular clustering 
and their evolution in the CDM model\thanks{
Based on data collected at Subaru Telescope, which is operated by 
the National Astronomical Observatory of Japan.}}

\author[T. Hamana et al.]
{Takashi Hamana$^{1}$, Toru Yamada$^{1}$, Masami Ouchi$^{2,3}$,
Ikuru Iwata$^{4}$, \and Tadayuki Kodama$^{1,5}$\\
$^1$ National Astronomical Observatory of Japan, 
Mitaka, Tokyo 181-8588, Japan\\
$^2$ Space Telescope Science Institute, 3700 San Martin Drive, 
Baltimore, MD 21218, USA\\
$^3$ Hubble Fellow\\
$^4$ Okayama Attrophysical Observatory, 
National Astronomical Observatory of Japan, Kamogata, Okayama 719-0232, Japan\\
$^5$ European Southern Observatory, Karl-Schwarzschild-Str. 2,
D-85748, Garching, Germany}

\date{Accepted ******; Received ******; in original form 2005 August 25}
\pagerange{\pageref{firstpage}--\pageref{lastpage}} 
\pubyear{2005} 
\volume{000}

\begin{document}

\label{firstpage}
\maketitle

\begin{abstract}
We examine masses of hosting haloes of two photometrically-selected 
high-$z$ galaxy samples: 
the old passively-evolving galaxies (OPEGs) at $z\sim 1$ 
and Lyman Break Galaxies (LBGs) at $z\sim 4$ both taken from the 
Subaru/XMM-Newton Deep Survey (SXDS).
The large survey area of the SXDS (1 deg$^2$) allows us to measure the 
angular two-point correlation functions to a wide separation of $>10$ arcmin
with a good statistical quality.
We utilize the halo model prescription for estimating characteristic masses 
of hosting haloes from the measured large-scale clustering amplitudes.
It is found that the hosting halo mass positively correlates with
the luminosity of galaxies.
Then, adopting the extended Press-Schechter model (EPS), we compute the 
predictions for the mass evolution of the hosting haloes in the
framework of the cold dark matter (CDM) cosmology
in order to make an evolutionary link between the two galaxy samples at
different redshifts and to identify their present-day descendants by
letting their haloes evolve forward in time.
It is found that, in the view of the mass evolution of hosting 
haloes in the CDM model, bright ($i'\la i'_\ast+1$) LBGs are consistent 
with being the progenitor of the OPEGs, whereas it is less likely that 
the LBG population, as a whole, have evolved into the OPEG population.
It is also found that the present-day descendants of both the bright
LBGs and OPEGs are likely to be located in massive 
systems such as groups of galaxies or clusters of galaxies.
Finally, we estimate the hosting halo mass of local early-type galaxy 
samples from the 2dF and SDSS based on the halo model and
it turns out that their expected characteristic mass of hosting
haloes is in good agreement with the EPS predictions for the
descendant's mass of both the bright LBGs and OPEGs.
\end{abstract}


\section{Introduction}
Galaxies consist of two major ingredients, namely, the baryonic and 
dark matter.
There are pieces of observational evidence that properties and 
evolution of the baryonic component are not independent of 
those of the dark matter, but are closely correlated.
For instance, early/late type galaxies dominate the 
bright/faint part of the galaxy luminosity function, 
which suggests that the star formation history and/or
morphology are related to the total mass of the system,
provided that the luminosity is basically proportional to the 
total mass.
Another example is the morphology-density relation (Dressler 1980;
Postman \& Geller 1984), which may indicate the possible influence
of the mass of hosting halo on the galaxy formation on group/cluster
scale as well. 
It follows from these pieces of evidence that the mass of the hosting
halo could be one of the most fundamental quantities in the galaxy
formation.
Therefore, in order to understand 
the galaxy formation, the evolution of baryonic and dark matter 
should not be treated as independent processes but should be 
investigated in a unified way.

The baryonic contents (stars and gas) of galaxies have been well
studied by both photometric and spectroscopic observations and
have been extensively analyzed with a help of well developed
theoretical tools such as stellar population synthesis models
(e.g., Bruzual \& Charlot 2003).
On the other hand, the dark matter component of hosting halos
has been less studied observationally due to the technical
difficulties in probing it since it requires dynamical
tracers or gravitational lensing.  Theoretical models
for the evolution of dark matter, however, have been well developed
with numerical simulations and analytical models such as
Press-Schechter prescriptions (Press \& Schechter 1974).

It has been known that a large-scale clustering amplitude
of galaxies provides a unique way of estimating their hosting halo mass 
in a statistical manner.
This makes use of a finding from theoretical studies of 
dark-matter structure formation in the cold dark matter (CDM) model 
that a clustering amplitude of dark matter haloes depends monotonically 
and strongly on the halo mass (Mo \& White 1996). 
By measuring the clustering amplitude of galaxy populations selected from 
a wide-field survey, we can relate properties of the populations such 
as stellar mass and/or star formation rate with the mass of their 
hosting halo (e.g., Giavalisco \& Dickinson 2001; Adelberger et al.~2005). 
It, in turn, enables us to explore an ancestor--descendant connection
of different galaxy populations at various redshifts from the viewpoints 
of both the star formation and the dark matter assembly 
(Moustakas \& Somerville 2002; Hamana et al.~2004; Ouchi et al 2004b).

This approach is exactly the one we take, in what follows, to explore
the evolution of the old passively-evolving galaxies (OPEGs) at
$z\sim 1$ and Lyman Break Galaxies (LBGs) at $z\sim 4$.
The OPEGs are selected by optical colour criteria
(see \S \ref{sec:data-opeg}) being consistent with a passively
evolving galaxy with the major star-formation epoch at $z > 2$.
Since the age of the universe at $z=2$ is only $\sim 2.3h^{-1}$Gyrs,
progenitors of the OPEGs must have experienced an active star formation 
phase at some high redshifts of $z>2$.
The LBGs, which are in general actively star forming galaxies at high redshifts,
are naturally a strong candidate of a major population of the
progenitor of such OPEGs.
While there is a fraction of dusty star-forming galaxies or red old
galaxies with little star formation, which are not selected by the LBG
criteria at $z\sim$2--2.5 (e.g., Franx et al. 2003; Reddy et al.
2005), LBGs are still most populous among them.
In addition, the OPEGs are considered as strong candidates for
progenitors of the present-day early-type galaxies since the simple
passive evolution in their stellar populations can easily link
between them.
We examine relations between the two galaxy populations in terms of 
the evolution of hosting dark matter haloes. 
Our primary question addressed in this paper is 
{\it ``supposing the LBGs are a major progenitor of the OPEGs,
are their hosting halo masses compatible 
with the mass evolution of dark matter haloes in the CDM structure 
formation model ?''}
Also, we discuss possible present-day descendants of
those high-$z$ galaxies from the viewpoint of the masses of hosting haloes.

We determine the clustering amplitude at large scales with
angular two-point correlation functions of the OPEGs
and LBGs measured, with a good accuracy, from a wide field (1 deg$^2$) 
deep imaging data-set of Subaru/XMM-Newton Deep Survey (SXDS).
Multi-band colour selection techniques successfully isolate those 
galaxies with well calibrated redshift selection functions, which
enable us to compute an accurate prediction for the corresponding dark 
matter angular correlation function via the Limber's projection. 
This large data-set and selection techniques allow us to compute a 
large-scale galaxy bias in a robust manner.
Comparing the measured large-scale bias with the
ones predicted by the halo model, we place a limit on 
the hosting halo mass of galaxy populations.
Here we adopt an empirically parameterized model 
for the halo occupation function (HOF) which describes statistical 
relations between galaxies and their hosting haloes.
Then, utilizing the extended Press-Schechter (EPS) prescriptions,
we examine the evolution of halo mass in the framework of the CDM model.
In this way, we compare hosting halo masses of each galaxy population
at different redshift, and explore their connection.

The outline of this paper is as follows.
Section 2 describes models and basic equations.
Section 3 summarizes observational data that are used to 
place a constraint on the hosting halo mass.
Results are presented in section 4.
Finally, section 5 is devoted to a summary and discussion.

Throughout this paper, we adopt a flat $\Lambda$CDM cosmology with 
the matter density $\Omega_{\rm m}=0.3$, the cosmological constant
$\Omega_\Lambda=0.7$, the Hubble constant 
$H_0=100 h$ km~s$^{-1}$~Mpc$^{-1}$ with  $h=0.7$, and the normalization of 
the matter power spectrum $\sigma_8=0.9$.
We adopt the fitting function of the CDM power spectrum of Bardeen et al.
(1986).
We present magnitudes in the AB system.

\section{Models}
\label{sec:models}

\subsection{Dark matter angular correlation function}
\label{sec:DMACF}

We quantify a clustering amplitude of a population of galaxies by
comparing their angular two-point correlation function with the 
corresponding dark matter correlation function.
Let $q(z)$ be a normalized redshift selection function of a population of
galaxies being considered, the dark matter angular two-point correlation 
function is computed from the dark matter power spectrum ($P_{\rm DM}$)
via the Limber projection 
(see e.g, chapter 2 of Bartelmann \& Schneider 2001):
\begin{equation}
\label{eq:limbers}
\omega_{DM}(\theta) = \int dr~ q^2(r) \int {{dk}\over {2 \pi}}
k~ P_{\rm DM} (k,r)~ J_0[f_K(r)\theta k],
\end{equation}
where $J_0(x)$ is the zeroth-order Bessel function of the first kind.  
For the spatially flat cosmology ($\Omega_{\rm m}+\Omega_\Lambda=1$) as 
we consider throughout the present paper, the radial function $f_K(r)$ 
is equivalent to $r$, and $r=r(z)$ is the radial comoving distance given by
$r(z) = c/H_0 \int_0^z dz'~[\Omega_{\rm m}(1+z')^3 + \Omega_\Lambda]^{-1/2}$.
We use the nonlinear fitting function of the CDM power spectrum by 
Peacock \& Dodds (1996).

\subsection{Halo model}
\label{sec:halomodel}

We utilize the halo model approach for estimating a characteristic mass of
hosting haloes from the measured clustering amplitude of galaxies.
Here, we summarize several expressions which are most relevant to the current 
analysis.
See Hamana et al.~(2004; and references therein) for details of the halo 
model.

We adopt a simple parametric form for the average number of a given galaxy 
population as a function of the hosting halo mass:
\begin{equation}
\label{eq:Ng}
  N_{\rm g}(M)=
\begin{cases}
{(M/M_1)^\alpha & ($M>M_{min}$)  \cr 
0 & ($M<M_{min}$) } 
\end{cases} ,
\end{equation}
which is characterized by the three parameters:
The minimum mass of haloes which host the population of galaxies 
($M_{min}$), a normalization parameter
which can be interpreted as the critical mass above which haloes typically
host more than one galaxy ($M_1$),
and the power-law index of the
mass dependence of the galaxy occupation number ($\alpha$).  
For the dependences of the HOF parameters on the shape of the two-point 
correlation function, see Hamana et al.~(2004; and references therein).

We introduce the average mass of hosting halo 
(weighted by the number of member galaxy) by
\begin{equation}
\label{eq:average-mass}
\langle M_{\rm host} \rangle = 
\frac{\int dM~ M~ N_{\rm g}(M) n_{\rm halo} (M)}
{\int dM~ N_{\rm g}(M) n_{\rm halo} (M)},
\end{equation}
where $n_{\rm halo}(M)$ denotes the halo mass function for which we
adopt the fitting function of Sheth \& Tormen (1999)

Since the halo model assumes the linear halo bias (Mo \& White 1996)
and $N_{\rm g}(M)$ solely depends on the halo mass, the galaxy bias 
on large scales (the scales larger than the virial radius of hosting haloes) 
is given by the galaxy number weighted halo bias:
\begin{equation}
\label{eq:bias}
b_{g,L}= 
\frac
{\int dM~ b_{\rm halo}(M) N_{\rm g}(M) n_{\rm halo}(M) }
{\int dM~N_{\rm g}(M) n_{\rm halo}(M) },
\end{equation}
where $b_{\rm halo}(M)$ is the halo bias, for which
we adopt the fitting function of Sheth \& Tormen (1999).
Note that $b_{g,L}$ does not depend on $M_1$ but only on 
$M_{min}$ and $\alpha$.

Since galaxies considered in the following sections
are distributed over a redshift interval, we take into account 
the redshift evolution of a quantity $X(z)$ (represents for $b_{g,L}^2$
or $\langle M_{\rm host} \rangle$) 
by computing its redshift average:
\begin{equation}
\label{eq:nga}
X \equiv \frac{\int dz~[dV/dz]~q(z)^k X(z)}
{\int dz~[dV/dz]~q(z)^k},
\end{equation}
where $k=1$ for $\langle M_{\rm host} \rangle$
while $k=2$ for $b_{g,L}^2$, and
$dV/dz$ denotes the comoving volume element per unit solid angle:
$dV/dz = c/H_0~r^2~[\Omega_{\rm m} (1 + z)^3 +\Omega_\Lambda]^{-1/2}$,
again for the spatially flat cosmology.

\subsection{Extended Press-Schechter model}
\label{sec:EPS}

The extended Press-Schechter (EPS) formalism was developed by 
Bond et al.~(1991), Bower (1991) and Lacey \& Cole (1993).
Since the EPS model provides a way to treat halo mergers, 
which play an important role in the structure formation 
in the hierarchical CDM model, it has been widely applied 
to analytical and semi-analytical studies of the structure 
formation.
We utilize the EPS model for making a statistical estimate of 
growth of halo masses.
A key expression for this is the conditional probability 
$P_2(M_{t2},z_2 | M_{t1},z_1)$ that a material in a halo of 
mass $M_{t1}$ at $z_1$ will be in a halo of mass $M_{t2}$ 
($M_{t2}>M_{t1}$) at a later redshift $z_2$, leading to an 
expression for the conditional mass function 
$n_2(M_{t2},z_2 | M_{t1},z_1)$.

Here we summarize only expressions which are directly relevant to 
our analysis, see above references for their derivation.
Let $\delta_c$ and $\sigma$ be the critical density threshold for a
spherical perturbation to collapse and the RMS density fluctuation
smoothed over a region enclosing a mass $M$, respectively, 
the conditional probability is 
\begin{eqnarray}
\label{p2}
\lefteqn{P_2(M_{t2},z_2 | M_{t1},z_1) dM_{t2}} \nonumber \\
& = & {1 \over {\sqrt{2 \pi}}}  
{{\delta_{c2} (\delta_{c1} - \delta_{c2})} \over \delta_{c1}}
\left[ {{\sigma_1^2} \over {\sigma_2^2 (\sigma_1^2 -\sigma_2^2)}} 
\right]^{3/2} \nonumber \\
&&  \times
\exp \left[ 
-{{(\sigma_2^2 \delta_{c1} - \sigma_1^2 \delta_{c2})^2}
\over
{ 2 \sigma_1^2 \sigma_2^2 (\sigma_1^2 - \sigma_2^2)}}
\right]
\left| {{d\sigma_2^2} \over {d M_{t2}}} \right|
 dM_{t2},
\end{eqnarray}
where the subscripts 1 and 2 stand for epochs $z_1$ and $z_2$,
respectively.
The conditional mass function is given from this by
\begin{equation}
\label{pn2}
n_2(M_{t2},z_2 | M_{t1},z_1) dM_{t2} \propto
{1 \over {M_{t2}}}  P_2(M_{t2},z_2 | M_{t1},z_1) dM_{t2} . 
\end{equation}

\section{Data}
\label{sec:data}

We use two samples of photometrically-selected galaxies from 
$B$, $R$, $i'$ and $z'$ imaging data of the 
Subaru/XMM-Newton Deep Survey\footnote{See 
http://www.naoj.org/Science/SubaruProject/SDS/ 
for details of the SXDS project.} (SXDS).
The limiting magnitudes for the 3$\sigma$ detection of an 
object in a 2 arcsec diameter aperture are 
$B \simeq 28.3$, $R \simeq 27.6$, $i'\simeq 27.5$ and
$z'\simeq 26.5$ (Furusawa et al. in preparation).
The seeing size of those images is $\sim 0.8\arcsec$.
We avoid the edges of the field and the area affected by bright sources.
The final effective area used in this paper is $\sim 1$ deg$^2$.

Below we summarize basic properties of the two galaxy samples, 
and present angular two-point correlation functions
of those galaxies.

\subsection{Old passively-evolving galaxies at $z \sim 1$}
\label{sec:data-opeg}

The photometric properties of the old passively-evolving
galaxies (OPEGs) in the SXDS fields are presented in
Kodama et al. (2004) and Yamada et al. (2005).
For the sample selection of OPEGs, we follow the definition of
Yamada et al. (2005) which imposes the two criteria:
$0.8 < i'-z' < 1.2 $ and 
$-0.05 z' + 3.01 < R -z' < -0.03 z' +2.49$,
on the $z'$-band selected catalog.
This criteria effectively isolate the galaxies with star-formation
epoch greater than $z_f > 2$ and located at $0.9 < z < 1.1$.
This photometrically-selected sample is composed of 
4,118 OPEG candidates with $z'<25.0$
distributed in the effective area of 1.03 deg$^2$.
It is found from the spectroscopic observations for 
the 93 OPEG candidates with $19 < z' < 22$ that 
the 73 objects lie between $z=0.87$ and 1.12, and the 4 objects 
lie outside of the redshift interval, and the remaining 16 objects 
have undetermined redshift because of no usable feature in the spectrum 
or a poor S/N.
Figure 8 of Yamada et al. (2005) shows the redshift distribution 
function of OPEGs obtained from the spectroscopically-identified objects.
The contamination fraction is estimated to be between
0.05 (=4/[73+4]) and 0.22 (=20/93). 
We assume the contamination fraction of $f_C=0.1$ in the following 
analyses.
The luminosity function is well fitted by the Schechter function
with $\phi_\ast = (4.26\pm 0.42)\times 10^{-3}h^3$Mpc$^{-3}$, 
$\alpha_{LF}=-0.67\pm 0.07$ and $M_{B*}=-21.38 \pm 0.10$ 
(Yamada et al. 2005).
The latter magnitude corresponds to $z'_\ast =21.8$ at $z=1$.

\begin{figure}
\begin{center}
\begin{minipage}{8.4cm}
\epsfxsize=8.4cm 
\epsffile{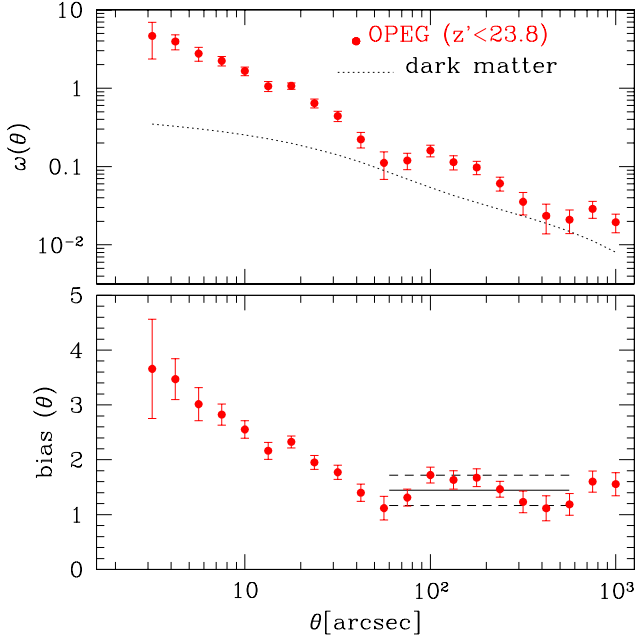}
\end{minipage}
\end{center}
\caption{{\it Upper panel}: Filled circles with error bars show 
the angular two-point correlation function for OPEGs with 
$z'<23.8(=z'_\ast+2)$. 
Note that the plotted correlation function has been corrected for
the contamination (see text). 
The dotted line shows the CDM model prediction 
for the dark matter angular two-point correlation function 
computed via equation (\ref{eq:limbers}), where the non-linear 
fitting function for the CDM power spectrum by Peacock \& Dodds (1996)
is used.
{\it Lower panel}: The corresponding galaxy bias defined by equation 
(\ref{eq:bias_g}) is plotted.
The horizontal solid line shows the large-scale bias factor 
computed averaging the bias over an interval $1\arcmin < \theta <10\arcmin$,
and the dashed lines show its 1-$\sigma$ error.}
\label{fig:acf-OPEG}
\end{figure}

The angular two-point correlation functions are 
computed using the pair-count estimator formulated
by Landy \& Szalay (1993):
$
\omega_g(\theta)= [DD(\theta)-2DR(\theta)+RR(\theta)]/RR(\theta).
$
In so doing, we distribute the same number of random samples with 
the same geometrical constraint as of the data sample.
We repeat 100 random re-samplings, and the mean and RMS among
the 100 measurements are taken as the mean correlation signal
and 1-$\sigma$ error.
The effect of the contamination on the two-point correlation function
is corrected by multiplying by a factor of $1/(1-f_C)^2$, where the 
contaminants are assumed to be randomly distributed.
A measured angular correlation function is plotted in the top panel 
of Figure \ref{fig:acf-OPEG} together with the dark-matter angular 
correlation function computed with the same redshift selection function.
Note that the integral constraint (Groth \& Peebles 1977)
is estimated to be 
$\sim 7 \times 10^{-3}$ for a single power-law model 
[$\omega_g(\theta) \propto \theta^\beta$] with 
the power-law index of $\beta=-1$, we may thus ignore an effect of 
a finite field size as far as $\theta < 10\arcmin$ is concerned. 

We define the galaxy bias by
\begin{equation}
\label{eq:bias_g}
b(\theta)=\sqrt{{\omega_g (\theta)} \over {\omega_{DM}(\theta)}},
\end{equation}
and is plotted in the bottom panel of Figure \ref{fig:acf-OPEG}.
As is shown there, on scales below $\sim 1$ arcmin, the bias 
decreases with the separation, while on larger scales it flattens.
The comoving angular length of the transition scale ($\sim$1\arcmin) 
is $\sim 0.7h^{-1}$Mpc (at $z=1$) which corresponds to the virial 
radius of the 
halo with the mass $\sim 2\times 10^{13}h^{-1}M_\odot$ at $z=1$.
This mass coincides with the characteristic mass of the hosting halo
($\langle M_{\rm host} \rangle$) predicted by the halo model 
analysis in the next section.
It may follow from this that the shape of the bias function
is basically understood by the standard halo model picture:
The small-scale clustering arises from galaxy pairs located in the 
same halo, while the large-scale clustering arises from galaxy 
pairs located in two different haloes.

\begin{figure}
\begin{center}
\begin{minipage}{8.4cm}
\epsfxsize=8.4cm 
\epsffile{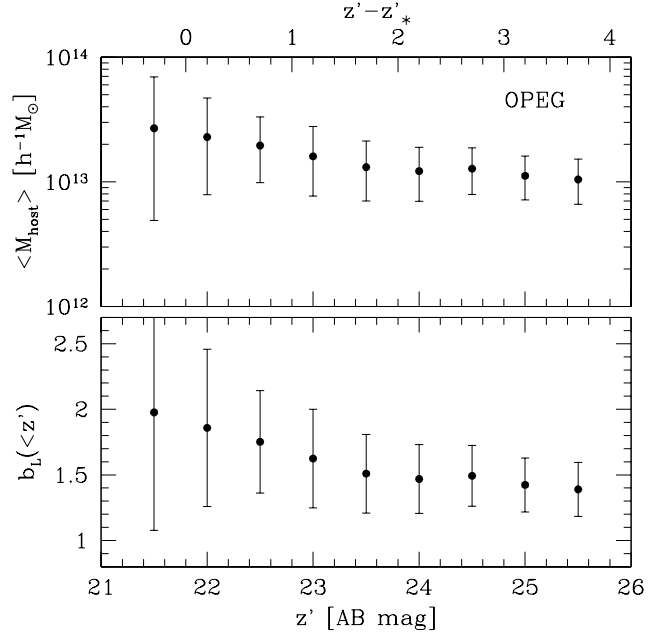}
\end{minipage}
\end{center}
\caption{{\it Lower panel}: The large-scale bias factor 
(see test for the definition) for the OPEGs as a function of
limiting magnitude.
{\it Upper panel}: $\langle M_{\rm host} \rangle$ computed from
the halo model (with a fixed $\alpha = 0.8$) for the corresponding 
large-scale bias factor (see \S \ref{sec:results} for details).
The upper abscissa axis indicates the magnitude difference from 
the $z'_\ast$ value.}
\label{fig:bL-OPEG}
\end{figure}

Since the measured bias flattens on large-scales as expected by 
the halo model, we define the large-scale bias factor ($b_L$) 
as an averaged bias over $1\arcmin < \theta < 10 \arcmin$ 
(illustrated in the bottom panel of Figure \ref{fig:acf-OPEG}).
We compute the large-scale bias factor for OPEG samples
with different limiting magnitudes and show in the bottom panel 
of Figure \ref{fig:bL-OPEG}.
Broadly speaking, the large-scale bias factor is a decreasing function 
of the magnitude, though the significance is not high due to 
large error bars.

Let us devote, in passing, a little space to compare our measurement with
previous studies.
We derive the comoving correlation length, $r_0$, from the power-law fitting
model of the angular two-point correlation function with the Limber
equation (Peebles 1980). The correlation length $r_0$ is 
the normalization of the spatial two-point correlation function, 
$\xi=(r/r_0)^{-\gamma}$, where $\gamma$ is related by $\gamma=\beta+1$.
We find $r_0 = 4.7 \pm 0.3h^{-1}$Mpc
and $5.7 \pm 0.2 h^{-1}$Mpc
with the best-fit beta ($\beta=1.1$) and
the fixed beta ($\beta \equiv 0.8$), respectively.
This is consistent with $r_0 = 5.0\pm 0.3 h^{-1}$Mpc obtained from 
$(R-I)$ colour selected red galaxies ($I=18-24$) at $z\sim 0.85$
most of which have early-type
spectra (Coil et al. 2004).
In contrast, it is much smaller than $r_0 = (8-13)h^{-1}$Mpc obtained from 
(optical$-$NIR) colour selected ``Extremely Red Objects'' (EROs) 
at $1 \la z \la 2$ (Daddi et al 2001; McCarthy et al. 2001; Firth et al.
2002; Roche et al. 2002; Miyazaki et al. 2003).
The luminosity segregation of clustering may not explain the
difference of $r_0$, 
since, if we assume an early-type spectrum,
the limiting magnitudes of their ERO searches 
reach to a comparable depth to our OPEG sample ($z'<23.8$).
The reason for the significant difference in $r_0$
is not clear.  Coil et al. (2004) suggest that
their red galaxies selected in optical bands may be
a less-extreme version of EROs.
Another possible reason would be an intrinsic evolution
in the clustering strength due to the difference in redshift
ranges of the samples.
While our sample and Coil et al.'s (2004) sample are both
limited to the small redshift range at $z \sim 1$,
other ``$R-K$" or ``$I-K$" EROs samples have much wider range
in redshift up to $z\sim2$
The clustering strength of the red galaxies can evolve strongly
between $z\sim2$ and 1.
Furthermore, the SXDS survey probes much larger volume than
the other surveys and the effect of field-to-field variation
is expected to relatively small.
Further investigating into the origin of the difference is, however,
beyond the scope of this paper, and we leave it for a future work.

\subsection{Lyman break galaxies at $z \sim 4$}
\label{sec:data-LBG}

\begin{figure}
\begin{center}
\begin{minipage}{8.4cm}
\epsfxsize=8.4cm 
\epsffile{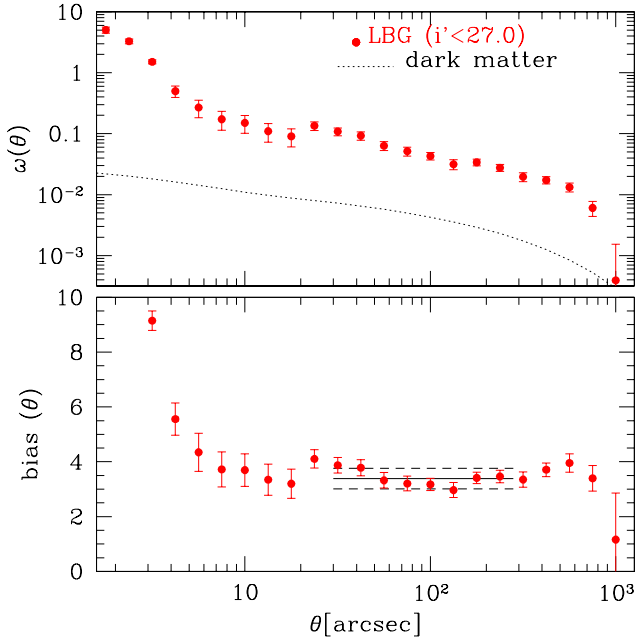}
\end{minipage}
\end{center}
\caption{Same as Figure \ref{fig:acf-OPEG} but for the LBGs with 
$i'<27.0(=i'_\ast+2)$.}
\label{fig:acf-LBG}
\end{figure}

\begin{figure}
\begin{center}
\begin{minipage}{8.4cm}
\epsfxsize=8.4cm 
\epsffile{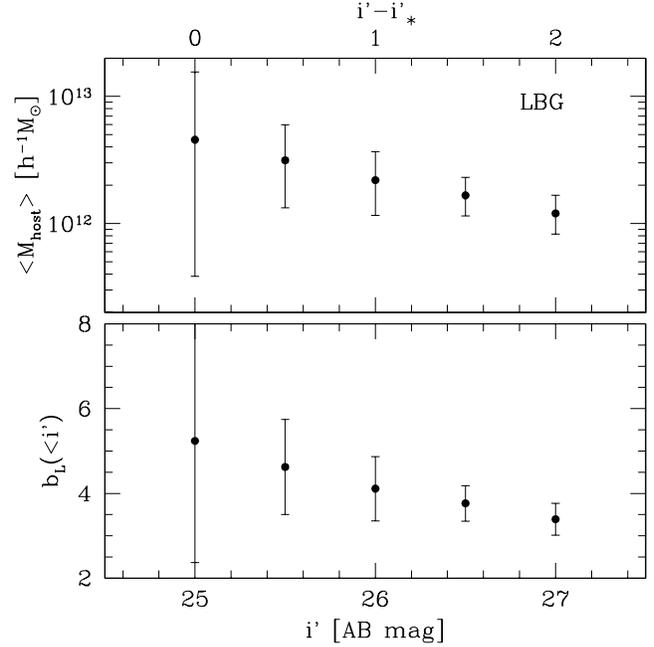}
\end{minipage}
\end{center}
\caption{Same as Figure \ref{fig:bL-OPEG} but for the LBGs.}
\label{fig:bL-LBG}
\end{figure}

The sample selection and clustering properties of the 
Lyman break galaxies (LBGs) at $z\sim 4$ are described in 
Ouchi et al. (2005).
Briefly, the LBGs are selected from $i'$-band selected catalog 
by three criteria: $B-R > 1.2$, $R- i'< 0.7 $ 
and $B-R > 1.6 (R-i')+1.9$.
The number counts of this sample is given in Table 1 of 
Ouchi et al. (2005).
The redshift distribution function has the mean of $z\sim 4$
and width of $\Delta z \sim 0.5$ as is shown in the top panels of 
Figure 12 of Ouchi et al. (2004a).
It is found from the spectroscopic follow-up observations for 
the 63 photometrically-selected LBGs that 
the 60 objects lie between $z=3.5$ and 4.5 (Ouchi et al. 2005
and references therein).
The contamination fraction is thus estimated to be 
$f_C =0.05 (=[63-60]/63)$.
Ouchi et al. (2004a) computed the luminosity function of the 
LBGs selected from the Subaru Deep field with the same colour selection
criteria and have found $i'_\ast=25.0\pm 0.1$.

The angular two-point correlation functions of the LBGs are computed in 
the same procedure as was done for the OPEGs and are plotted in the top 
panel of Figure \ref{fig:acf-LBG}.
The integral constraint is estimated to be 
$\sim 7 \times 10^{-3}$ for a single power-law model with 
the power-law index of $\beta=-0.8$, we may thus ignore an
effect of a finite field size as far as $\theta < 5\arcmin$
is concerned.
The galaxy bias defined by equation (\ref{eq:bias_g}) is 
plotted in the bottom panel. 
A transition from the small-scale decreasing part to 
the large-scale flat part is found at the scale of $\sim$10 arcsec 
(see Ouchi et al. 2005 for further detail discussions on the shape of 
the bias function. 
Note that they do not correct for the effect of the contamination
because of unknown clustering property of the contaminants,
for which we assume the random distribution.  As a consequence,
their measured biases are smaller than ours by about 5 percents).
This is transformed to the comoving angular scale of
$\sim$0.2$h^{-1}$Mpc at $z=4$, which corresponds to the virial radius
of the halo with the mass $\sim 1\times 10^{12}h^{-1}M_\odot$.
This mass, again, coincides with the characteristic mass of hosting haloes
($\langle M_{\rm host} \rangle$) predicted by the halo model 
analysis in the next section.
Thus the behavior of the bias is understood by the standard 
halo model picture.

The large-scale bias factor is computed in the same manner as 
was done for the OPEGs except for using a different separation range 
of $0.5\arcmin < \theta < 5\arcmin$
(illustrated in the bottom panel of Figure \ref{fig:acf-LBG}).
This bias factor is calculated as a function of limiting magnitudes
and is plotted in the bottom panel of Figure \ref{fig:bL-LBG}.
As was reported by Ouchi et al. (2005), a trend that the large-scale
bias factor decreases with the luminosity is observed.

\section{Results}
\label{sec:results}

\begin{figure}
\begin{center}
\begin{minipage}{8.4cm}
\epsfxsize=8.4cm
\epsffile{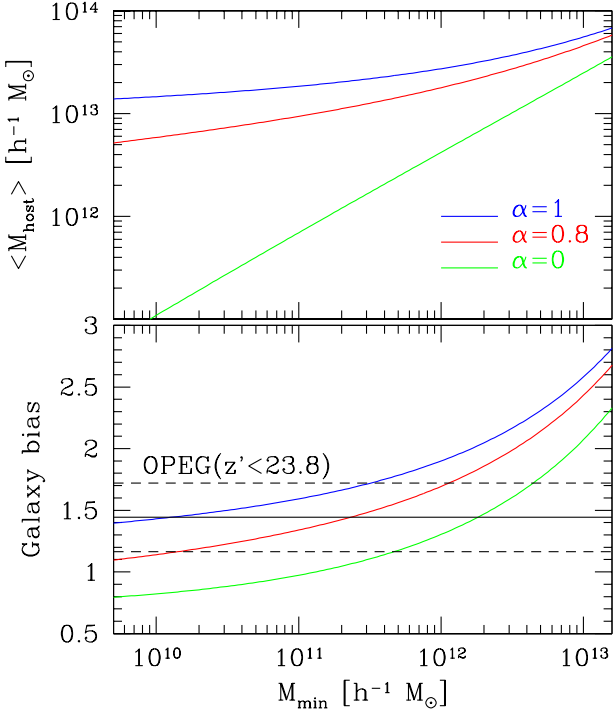}
\end{minipage}
\end{center}
\caption{Halo model predictions for OPEGs.
{\it Lower panel}: The large-scale bias $b_{g,L}$ defined by equation 
(\ref{eq:bias}) for three values of $\alpha$ (1.0, 0.8 and 0 from 
upper to lower). 
The horizontal lines show the measured large-scale bias factor for
the OPEG sample with $z'<23.8(=z'_\ast+2)$, the solid and dashed line 
represent the mean and 1-$\sigma$ range, respectively.
{\it Upper panel}: The mean mass of hosting haloes defined by equation
(\ref{eq:average-mass}) for $\alpha= 1.0$, 0.8 and 0 from upper to lower.}
\label{fig:halo-OPEG}
\end{figure}

We define fiducial subsamples of the two galaxy populations for the 
later analyses by setting 
the magnitude limit to 2 magnitude fainter than the $m_\ast$ value
($z'_\ast=21.8$ for the OPEGs and $i'_\ast=25.0$ for the LBGs).
This is chosen so that it is well deeper than the $m_\ast$ value 
and at the same time it is well brighter than the completeness limit of
our imaging data ($z' \simeq 26.5$ for the OPEGs and $i' \simeq 27.5$ for 
the LBGs).
The former is imposed so that the samples do not contain only the
brightest objects, which would not be representative of the whole
population.
The latter is imposed to reduce contaminations due to errors in the 
luminosities and colours.
We note that the number densities of the subsamples are 
$n_{\rm OPEG}(z'<23.8) = (4.7\pm 0.47)\times 10^{-3} h^3$Mpc$^{-3}$
and 
$n_{\rm LBG}(i'<27.0) = (1.1\pm 0.2)\times 10^{-2} h^3$Mpc$^{-3}$ 
for OPEGs and LBGs, respectively.
Therefore with this selection, 
the LBGs are as 2.3 times numerous as the OPEGs.
It should be noted that statistical properties 
(e.g., the number density and clustering amplitude) of a galaxy sample 
would be sensitive to the selection criteria
(e.g., the magnitude limit and colour selection).
Therefore, when one attempts to compare properties of two (or more) galaxy 
samples, it is necessary to properly define samples of galaxies 
with a quantity essential to their nature.
In our case, the luminosity would be the most relevant quantity among
(a few) controllable parameters we have in hand. 
Since the definition of our subsamples is based on
a somewhat arbitrary magnitude limit, we shall 
look into an effect of this definition by changing the magnitude limit 
for subsampling.

\subsection{Halo model analyses}

\begin{figure}
\begin{center}
\begin{minipage}{8.4cm}
\epsfxsize=8.4cm 
\epsffile{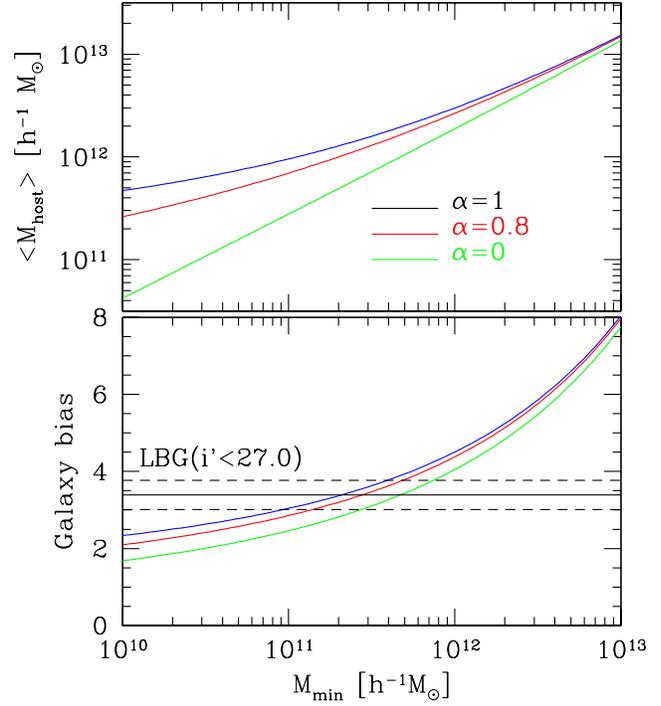}
\end{minipage}
\end{center}
\caption{Same as Figure \ref{fig:halo-OPEG} but for the LBGs.
Measured large-scale bias factor and number density shown by
the horizontal lines are taken from the sample $i'<27.0(=i'_\ast+2)$.}
\label{fig:halo-LBG}
\end{figure}

\begin{table}
\caption{Constraints on $\langle M_{\rm host} \rangle$ obtained
from the halo model analysis.
Unit of the mass is in [$h^{-1}M_\odot$].
$^{(1)}$ values of $\langle M_{\rm host} \rangle$ estimated from the 
mean of $b_L$. $^{(2)}$ the interval of $\langle M_{\rm host} \rangle$ 
from the 1-$\sigma$ range of $b_L$.}
\label{table:Mhalo}
\begin{tabular}{llcc}
\hline
&  & mean$^{(1)}$ & 1-$\sigma^{(2)}$ \\
\hline
OPEGs & $\alpha=1$ & $1.5\times 10^{13}$ & $9.6\times 10^{12}-2.2\times 10^{13}$\\
($z'<23.8$)& $\alpha=0.8$ & $1.2\times 10^{13}$ & $6.3\times 10^{12}-1.9\times 10^{13}$\\
& $\alpha=0$ & $6.6\times 10^{12}$ & $2.3\times 10^{12}-1.3\times 10^{13}$\\
%
LBGs & $\alpha=1$ & $1.3\times 10^{12}$ & $9.3\times 10^{11}-1.8\times 10^{12}$\\
($i'<27.0$)& $\alpha=0.8$ & $1.2\times 10^{12}$ & $8.3\times 10^{11}-1.7\times 10^{12}$\\
& $\alpha=0$ & $1.0\times 10^{12}$ & $6.4\times 10^{11}-1.5\times 10^{12}$\\
\hline
\end{tabular}
\end{table}

We estimate the hosting halo mass by comparing
the measured large-scale bias factor with the halo model prediction.
To do this, we proceed as follows: 
First, we compute the halo model prediction for the large-scale bias 
$b_{g,L}$ defined by equation 
(\ref{eq:bias}) for a given value of $\alpha$ as 
a function of $M_{min}$ (plotted in the lower panel of Figures 
\ref{fig:halo-OPEG} and \ref{fig:halo-LBG}).
Then, searching for an interval of $M_{min}$, where the predicted 
$b_{g,L}$ and the 1-$\sigma$ interval of the measured large-scale bias 
factor intersect, we have a constraint on $M_{min}$.
Since for a given $\alpha$, $\langle M_{\rm host} \rangle$ and $M_{min}$
have a one-to-one correspondence (see the upper panel of Figures 
\ref{fig:halo-OPEG} and \ref{fig:halo-LBG}), the constraint on
$M_{min}$ is immediately translated into the constraint on 
$\langle M_{\rm host} \rangle$.
We take three values of $\alpha = 1$, 0.8 and 0.
Here $\alpha < 1$ is preferred from semi-analytic models of the 
galaxy formation (e.g., Kravtsov et al. 2004),
as well as from the halo model analysis for LBGs (Hamana et al. 2004).
We take $\alpha =0.8$ as a fiducial value.
The case $\alpha=0$ corresponds to an extreme case where
every halo with a mass greater than $M_{min}$ always has one galaxy.

The constraints obtained for our fiducial galaxy samples 
($z'<23.8$ for the OPEGs and $i'<27.0$ for the LBGs) are summarized in 
Table \ref{table:Mhalo}.
It is important to notice that the change in $\alpha$ does not make a 
significant change in the preferred interval of 
$\langle M_{\rm host} \rangle$ 
except for the extreme case of $\alpha=0$ for the OPEGs.
Therefore , the constraint is not very sensitive to the
uncertainty in $\alpha$.

Let us look into how the characteristic hosting halo mass 
varies with the limiting magnitude for the sample selection.
The upper panels of Figures \ref{fig:bL-OPEG} and \ref{fig:bL-LBG}
show the constraint on $\langle M_{\rm host} \rangle$ as a function 
of the limiting magnitude.
In both galaxy populations, a trend of decreasing 
$\langle M_{\rm host} \rangle$ for a fainter limiting magnitude is observed.
These similar-looking trends, however, may arise from different physical
origins as explained below.
First, for the OPEGs, the observed $z'$-band corresponds approximately to
the rest-frame $B$-band.
Although the $B$-band luminosity is affected by on-going/recent
star formation, only a weak star formation activity would make the
colour of galaxy blue and pushes it outside of our OPEG 
selection criteria (Yamada et al. 2005).
Therefore the $B$-band luminosity is a good measure of the stellar 
mass of the OPEGs. 
Accordingly, the observed trend is considered as a result of a
correlation between the stellar mass and the hosting halo mass.
On the other hand, for the LBGs, the observed $i'$-band 
corresponds to $\sim$1500{\AA} in the rest-frame, where the 
luminosity is most sensitive to the star formation.
The LBGs are generally in an active star formation phase, 
also it was reported that the stellar mass of LBGs is poorly 
correlated with the UV luminosity (Shapley et al. 2001).
Therefore, the observed luminosity dependence of hosting halo
mass of LBGs may suggest a presence of a correlation between 
the hosting halo mass and the star formation activity.

\begin{figure}
\includegraphics[height=84mm,angle=-90]{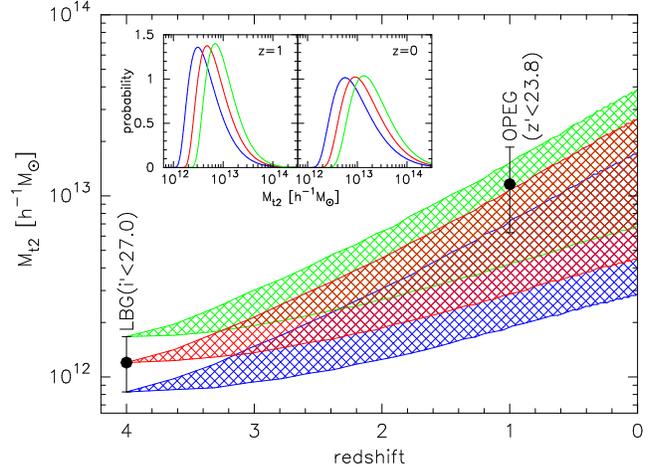}
\caption{Inserts show arbitrary normalized conditional mass functions
$n_2(M_{t2},z_2|M_{t1},z_1)$ for $z_2=1$ (left panel) and 0 (right panel),
where parameters of the earlier halo is taken from the limit obtained 
by the halo model analysis of the LBGs (with $\alpha=0.8$): 
$z_1=4$, $M_{t1}=1.2\times 10^{12}h^{-1}M_\odot$ (red), 
$8.3\times 10^{11}$ (blue), and $1.7\times 10^{12}$ (green).
These conditional mass functions are considered as the probability 
distribution function of the mass of the descendant.  
The main plot shows 68\% confidence intervals of $M_{t2}$ 
as a function of the redshift ($z_2$).
Three intervals shown by coloured hatches are for the same three $M_{t1}$ 
values as the inserts.
Filled circles with error bars show the limits of  
$\langle M_{\rm host} \rangle$ obtained from the halo model analysis 
(with $\alpha=0.8$) for the LBGs and OPEGs.}
\label{fig:eps-LBG}
\end{figure}

\subsection{EPS model analyses}

\begin{figure}
\includegraphics[height=84mm,angle=-90]{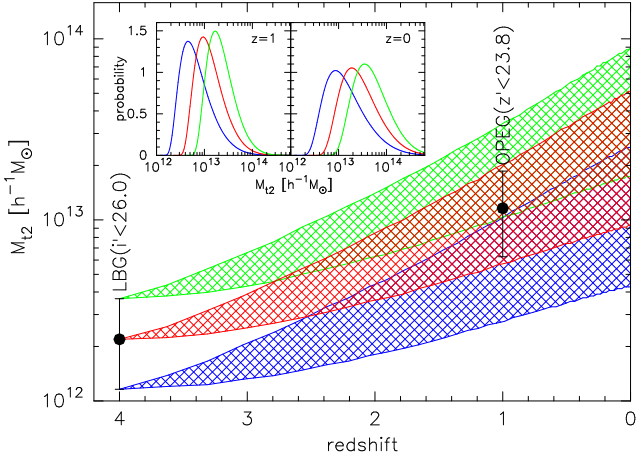}
\caption{Same as Figure \ref{fig:eps-LBG} but the LBGs sample is
selected with $i'<26.0$.
The limit of $\langle M_{\rm host} \rangle$ obtained from the halo 
model analysis with $\alpha=0.8$, which is taken by $M_{t1}$, is 
$M_{t1}=2.2\times 10^{12}h^{-1}M_\odot$ (red), 
$1.2\times 10^{12}$ (blue), and $3.7\times 10^{12}$ (green). }
\label{fig:eps-LBG26}
\end{figure}

We use the conditional mass function $n_2(M_{t2},z_2 | M_{t1},z_1)$ 
(equation \ref{pn2}) derived from the EPS formalism to predict the 
mass evolution of hosting haloes in the framework of the CDM cosmology.
For the parameters of the haloes at earlier epoch $z_1$,
we take the 1-$\sigma$ interval 
of $\langle M_{\rm host} \rangle$ obtained from the halo model analysis  
with $\alpha=0.8$ as a range of $M_{t1}$, and we set $z_1=4$ and $z_1=1$ 
for the LBGs and OPEGs, respectively.
Then for a certain later epoch $z_2$, we compute 
$n_2(M_{t2},z_2 | M_{t1},z_1)$ as a function of $M_{t2}$ (see inserts 
of Figures \ref{fig:eps-LBG}--\ref{fig:eps-OPEG} for examples).
Assuming that the mass assembly history of hosting haloes is not biased
toward any specific merging path, the conditional mass function 
can be regarded as the probability distribution function (PDF) of the mass 
of descendant haloes $M_{t2}$.
We define a 68\% confidence interval of $M_{t2}$ 
as the EPS model prediction for the mass of the descendant halo.

Such predictions are made for our LBGs and OPEGs samples.
and are shown in Figures 
\ref{fig:eps-LBG}--\ref{fig:eps-OPEG} as hatched regions.
Three tracks correspond to the three different values of
$M_{t1}$ taken from the mean and upper/lower 1-$\sigma$ values for 
the $\langle M_{\rm host} \rangle$
denoted by the symbols with error bars 
($\alpha=0.8$ cases in Table \ref{table:Mhalo}).
The inserts show the PDFs of $M_{t2}$ at $z=0$ (and $z=1$).
It is observed in the inserts that the PDFs have a wide spread, implying 
a wide variety of the mass assembly history of hosting haloes.
We note that since the PDFs are skewed toward a larger mass, 
the mean of the distribution is greater than the mode.

Let us look closely at each result.
First of all, Figure \ref{fig:eps-LBG} compares the expected
descendant mass of 
hosting haloes of the LBGs with the $\langle M_{\rm host} \rangle$ of OPEGs.
Both galaxy samples are selected with our fiducial magnitude limit of 
$m < m_\ast+2$.
As is evidently shown, the EPS predictions for the mass of the LBG descendant
are slightly smaller than the predicted mass range of the OPEGs, only the 
track from the upper limit of the LBG haloes is compatible with the OPEGs.

Is there a selection criterion for LBGs or for OPEGs which results in a
better agreement?
It is found from the upper panel of Fig. \ref{fig:bL-OPEG} that choosing 
a fainter limiting magnitude for the OPEG sample lowers the 
$\langle M_{\rm host} \rangle$ very little and can hardly solve the
incompatibility.
On the other hand, choosing a brighter limiting magnitude for the LBG sample 
raises the $\langle M_{\rm host} \rangle$ as shown in the upper panel of 
Figure \ref{fig:bL-LBG}.
Indeed, the galaxy sample selected with $i' < 26.0(=i'_\ast+1)$ results in a
very compatible halo mass with the $\langle M_{\rm host} \rangle$ of the OPEGs
(with $z'<23.8$) as shown in Figure \ref{fig:eps-LBG26}.
Therefore, in the view of the mass evolution of hosting haloes, 
it may be safely concluded that the bright ($i'\la i'_\ast+1$) LBGs are 
consistent with being the progenitor of the OPEGs, whereas it seems less 
likely that the LBGs population, as a whole, 
has evolved into the OPEG population.
Note that the corresponding number density is computed to be
$n_{\rm LBG}(i'<26.0) \simeq (3.8 \pm 0.3) \times 10^{-3} h^{3}$Mpc$^{-3}$,
closer to the number density of $n_{\rm OPEG}(z'<23.8).$

\subsection{Predictions for the present-day descendants}
\label{sec:presentday}

\begin{figure}
\includegraphics[height=84mm,angle=-90]{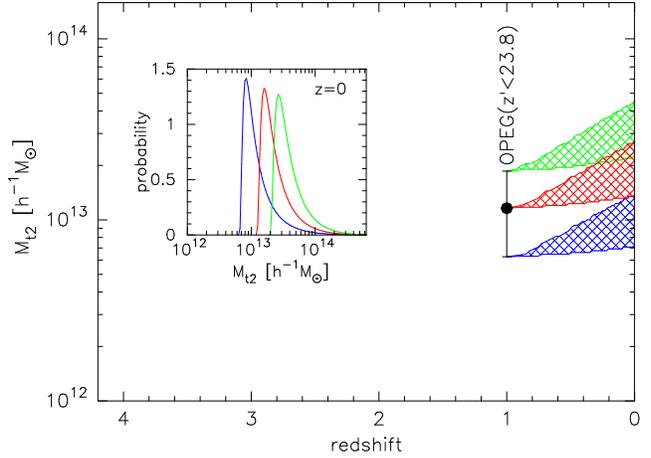}
\caption{Same as Figure \ref{fig:eps-LBG} but the limit for 
OPEGs obtained from the halo model analysis (with $\alpha=0.8$) 
is taken by parameters of the earlier halo:  $z_1=1$, 
$M_{t1}=1.2\times 10^{13}h^{-1}M_\odot$ (red), 
$6.3\times 10^{12}$ (blue), and $1.9\times 10^{14}$ (green).
In the insert, $z_2$ is taken by 0.}
\label{fig:eps-OPEG}
\end{figure}

Turn next to predictions for the present-day descendants of the two galaxy
populations plotted in Figures \ref{fig:eps-LBG}--\ref{fig:eps-OPEG}.
As is shown in the inserts of Figures \ref{fig:eps-LBG} and \ref{fig:eps-LBG26},
the PDF of the present-day descendants of the LBGs have a very broad spread 
due to the wide variety of the mass assembly history over $\sim 8h^{-1}$Gyrs.
The predicted mass range for the LBG sample with $i'<27.0$ is  
broadly consistent with that of the OPEGs, though the PDF of the LBGs extends 
to a smaller mass (say $M<10^{13}h^{-1}M_\odot$) where the PDF of the OPEGs 
has little probability.
On the other hand, the predicted mass range for the LBG sample with 
$i'<26.0$  agrees better with that of the OPEG descendants.
In this case, both the PDFs computed from the central 
$\langle M_{\rm host} \rangle$ value 
(the middle one of the three cases in the inserts) peak at 
$\sim 2 \times 10^{13} h^{-1} M_\odot$, which corresponds to the mass 
scale of groups of galaxies.
It is important to note that since the PDFs are skewed strongly toward 
a larger mass, a certain fraction of haloes is expected to 
evolved into more massive haloes with
$M_{t2} \ga 10^{14} h^{-1} M_\odot$ (see the inserts of 
Figures \ref{fig:eps-LBG26} and \ref{fig:eps-OPEG}).

\begin{figure}
\begin{center}
\begin{minipage}{8.4cm}
\epsfxsize=8.4cm 
\epsffile{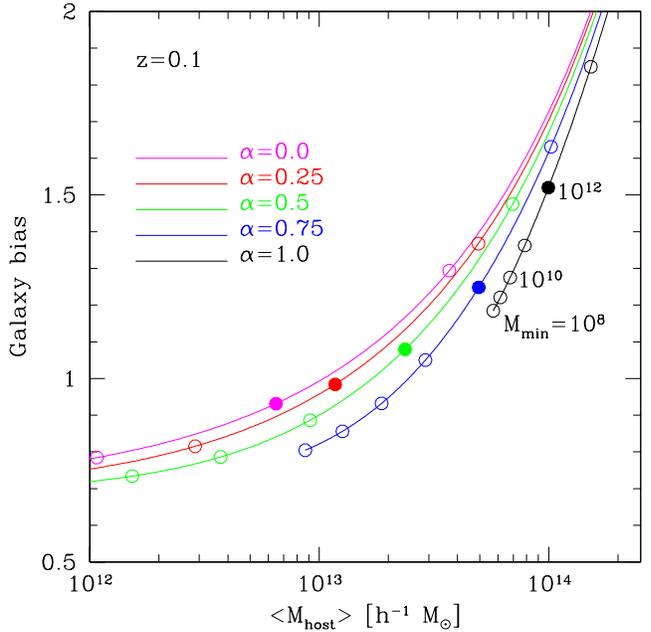}
\end{minipage}
\end{center}
\caption{The halo model prediction for the large-scale galaxy bias
$b_{g,L}$ as a function of the 
averaged halo mass $\langle M_{\rm host} \rangle$ at $z=0.1$.
Curves are for a different value of $\alpha$ 
($\alpha=0$, 0.25, 0.5, 0.75 and 1 from upper to lower).
Open circles indicate a corresponding value for $M_{min}$ 
in intervals of $\Delta \log M_{min}=1$ staring from 
$M_{min} = 10^8 h^{-1}M_\odot$ with 
$M_{min} = 10^{12} h^{-1}M_\odot$ being 
marked by the filled circles.}
\label{fig:z0.1}
\end{figure}

It is interesting to compare those EPS predictions
for the halo masses of the present-day descendants 
with local galaxy samples.
Since both the LBGs and OPEGs are frequently argued as the strong 
candidates for the progenitor of the present-day early-type galaxies, 
we take two clustering analyses of the early-type galaxy samples:
One from the the 2dF (Madgwick et al. 2003), 
and the other from the SDSS (Zehavi et al. 2002).
Madgwick et al. (2003) found 
$\sigma_8^{\rm NL} \simeq 1.1\pm0.1$ for ``passive'' galaxies,
of which the luminosity range is $-16.5 < M_{b_J} - 5 \log h < -22$ with 
$M_{b_J,\ast}=-19.7$.
Here $\sigma_8^{\rm NL}$ is the RMS of counts of galaxies in spheres of 
$8h^{-1}$Mpc radius and is derived from the best-fitting power-law model 
of the real-space correlation function $\xi=(r_0/8)^\gamma$ via the relation
$(\sigma_8^{\rm NL})^2 = J_2(\gamma) (r_0/8)^\gamma$ with
$J_2(\gamma) = 72/[(3-\gamma)(4-\gamma)(6-\gamma)2^\gamma]$ 
(Peebles 1980).
Since $\sigma_8^{\rm NL}$ of the dark matter 
is $\sigma_8^{\rm NL}\simeq 0.9$ at $z=0.1$ (the mean redshift of the 
galaxy samples considered here) for $\sigma_8=0.9$ 
(note that $\sigma_8$ is the linearly extrapolated value at $z=0$), 
the corresponding bias is found to be $b_{g,L}\simeq 1.2\pm0.1$.
Following the same procedure, we estimate $\sigma_8^{\rm NL}$ of 
galaxy samples from the SDSS.
Adopting the best-fitting power-law models given in 
Table 2 of Zehavi et al. (2002), we find $\sigma_8^{\rm NL}=1.2\pm 0.03$ 
for their both ``red'' (defined by $u-r>1.8$) and ``high concentration'' 
($c=r_{90}/r_{50}>2.7$) samples, giving $b_{g,L}\simeq 1.3\pm0.03$ 
for $z=0.1$ (the mean redshift of the samples).
The luminosity range of those galaxies is 
$-22 < M_r -5 \log h< -19$ with $M_{r,\ast}=-20.8$.
Comparing those bias values (i.e., $b_{g,L}=1.1-1.33$) 
with the halo model predictions plotted in Figure\ref{fig:z0.1}, we find 
$\langle M_{\rm halo} \rangle = (2-6)\times 10^{13}h^{-1}M_\odot$
for $0.25 \la \alpha \la 0.75$ (Note that as is shown Figure\ref{fig:z0.1}, 
the relation between the large-scale
bias and $\langle M_{\rm halo} \rangle$ only weakly depends on $\alpha$).
This is broadly consistent with the EPS predictions for the descendants
of the OPEGs (with $z'<23.8$ ) and also with that of the LBGs (with $i'<26.0$).
Therefore, we may conclude that, in the viewpoint of the mass evolution of 
hosting haloes in the CDM model, the OPEGs and the bright LBGs are consistent 
with being the progenitor of the present-day early-type galaxies.

\section{Summary and discussions}
\label{sec:summary}

We have analyzed the two photometrically-selected 
galaxy samples from the deep 
mult-band images of the SXDS: the LBGs at $z\sim 4$ and OPEGs at $z\sim 1$.
The contiguous large survey area of the SXDS enables us to measure the 
angular two-point correlation functions over a wide range of separations 
with a good statistical quality.
Comparing the large-scale clustering amplitude with the corresponding 
halo model predictions, we have estimated a characteristic mass of 
hosting haloes.
Then, adopting the EPS model, we have 
computed the predictions for the mass evolution of the hosting haloes 
to explore the likely descendants of those galaxy samples.
In particular, we have compared the predicted halo masses of these
two populations at different epochs to make a possible evolutionary
link.
We have also examined expected halo masses of the present-day descendants
of those two galaxy populations.

Our major findings are summarized as follows:
\begin{enumerate}
\item The measured bias functions (defined by equation \ref{eq:bias_g})
of both the OPEGs and LBGs consist of two parts: The decreasing small-scale
part and the flat large-scale part.
The transition scales between them are 1{\arcmin} and 10{\arcsec} 
for the OPEGs and LBGs, respectively.
Those scales agree well with the virial radii of haloes with the 
characteristic mass of hosting haloes predicted by the halo model analysis 
in \S \ref{sec:results}.
The shape of the bias function is thus understood by the standard 
halo model picture: The small-scale clustering arises from galaxy pairs 
located in the same halo, and the large-scale clustering arises from galaxy 
pairs located in two different haloes.
\item The conclusive measurement of the flat large-scale bias allows us to
safely define the large-scale bias factor.
Also, the large number of galaxies allows us to examine the luminosity 
dependence of the large-scale bias factor.
It is found that the large-scale bias factors of both OPEGs and LBGs 
positively correlates with the luminosity (see bottom panels of Figures 
\ref{fig:bL-OPEG} and \ref{fig:bL-LBG}).
\item Comparing the measured large-scale bias factors of the two 
galaxy samples with the halo model predictions, we estimate the 
characteristic mass of hosting haloes.
The merit of using the large-scale bias factor is two-fold:
First is concerning the measurement; 
since many bins are used to compute it, the measurement
is less sensitive to a statistical noise in each bin.
Second is concerning the halo model; 
the large-scale bias does not depend on $M_1$, and in addition,  
the resultant $\langle M_{\rm halo} \rangle$ is less sensitive to 
$\alpha$.
Therefore it provides with a reliable estimate of 
$\langle M_{\rm halo} \rangle$.
The predicted characteristic halo masses of both galaxy samples are 
found to be positively correlated with the luminosity.
For the OPEGs, this may arise from a 
correlation between the stellar mass and the hosting halo mass,
because the observed $z'$-band magnitude (rest-frame $B$-band)
is well correlated to the stellar mass for population like OPEGs
with little star formation activity.
On the other hand, for the LBGs, it may suggest a correlation 
between the hosting halo mass and the star formation activity, 
rather than the stellar mass.
This is because the observed $i'$-band 
corresponds to $\sim$1500{\AA} in the rest-frame, where the 
luminosity is most sensitive to the star formation, and in addition, 
the LBGs are generally in an active star formation phase.
\item Utilizing the EPS model, we compute the predictions for the halo 
mass distribution of the LBGs' descendants
at $z=1$ in the CDM cosmology, and then we 
compare it with the halo model prediction for the characteristic halo mass 
of the OPEGs.
It is found that, for our fiducial subsamples (the OPEGs with $z'<23.8$ 
and the LBGs with $i'<27.0$), the typical hosting halo mass of LBGs'
descendants is 
slightly smaller than the predicted mass of the OPEGs' hosting haloes.
It is also found that the brighter LBG subsample (with $i' \la 26.0$) 
is likely to evolve into the systems with halo mass compatible
to the predicted one of the OPEGs.
Therefore, we may conclude that, in the viewpoint of the mass evolution of 
hosting haloes in the framework of the CDM model, 
the bright ($i'\la i'_\ast+1$) LBGs are 
consistent with being the progenitor of the OPEGs.
Accordingly, it seems less likely that the LBGs population, as a whole, 
has evolved into the OPEG population.
\item We also compute predictions for halo masses of the present-day 
descendants of both the galaxy samples using the EPS model.
It is found that the predicted mass range for the LBG sample with 
$i'<27.0$ is  
slightly but systematically smaller than that of the OPEGs (with $z'<23.8$).
On the other hand, the prediction for the LBG sample with 
$i'<26.0$ agrees better with that of the OPEG descendants.
In the latter case, the peaks of the PDFs are located at 
$\sim 2 \times 10^{13} h^{-1} M_\odot$ and the tail of the PDFs extends 
to the mass range of $M>10^{14} h^{-1} M_\odot$.
Thus the present-day descendants of the bright LBGs and the OPEGs 
are likely to be located in massive systems such like groups of galaxies 
or clusters of galaxies.
We also estimate the characteristic halo mass of local early-type galaxy 
samples from the 2dF and SDSS with the halo model, and
it turns out that the predicted mass is in good 
agreement with the EPS predictions for the present-day descendant's mass 
of both the bright LBGs and OPEGs.
Therefore, it is concluded that, in the viewpoint of the mass evolution of 
hosting haloes in the CDM model, the OPEGs and bright LBGs are consistent 
with being the progenitor of the present-day early-type galaxies.
\end{enumerate}

One of the most interesting implications from the above findings is that
the possible halo mass dependence of the LBG's star formation history.
This is speculated from the above finding (iv) that the predicted descendant's 
halo mass of bright LBG subsample ($i'\la i'_\ast+1$) is found to be 
in very good agreement with the characteristic halo mass of the OPEGs,
whereas it seems less likely that the faint LBG population has 
evolved into the OPEG population.
The nature of the descendants of the faint LBGs at $z \simeq 1$ is not 
clear but is naively expected that they evolve into other populations than
old passively evolving galaxies with lower mass and bluer spectra.
It is important to notice that a little star formation at $z<2$ would be
enough to push the galaxy colour outside of the OPEG's colour criteria
(Yamada et al. 2005).
Therefore, the halo mass dependence of the epoch of truncation in star 
formation activity is one possibility of interpreting the finding (iv).
It is also found that the UV luminosity of the LBGs correlates with 
the hosting halo mass [the above finding (iii)].
This may be an additional evidence for the mass dependent star formation 
history, though a connection between the above two findings, (iii) and (iv), 
is not clear.
A possible scenario is that LBGs in more massive haloes have more active 
star formation at significantly high redshift such as $z\sim4$,
and they turn into the passive evolution phase earlier.
This scenario can be tested by performing the same clustering analysis
as presented in this paper progressively toward lower (and higher)
redshifts.  We might be able to see the transition at some point
where star formation activities in massive haloes are truncated and the
mass of the haloes that are hosting active star formation is being
shifted to lower mass as time progresses.
At the same time, on the theoretical side, we should explore
a possible physical mechanism that drives such mass dependent
star formation histories.

Before closing, it is important to note that we have {\it not} argued
that LBGs are the only path to become OPEGs at $z=1$ or the present-day
early-type galaxies.
In fact, it is likely that the ancestor--descendant connection is not 
a one-to-one correspondence, and some different high-$z$ galaxy
populations may have evolved into a similar low-$z$ population,
and vice versa.
For example, among the ancestors of the OPEGs at $z=1$, there may be
some haloes which have not collpased by $z=4$ and would be seen as
LBGs at some later epochs, as well as the objects which are not
UV luminous enough to be selected as LBGs at $z=4$ due to large
amount of dust extinction and/or older stellar ages.
This is one reason why we have not take into account the number density
of our galaxy samples when we examine their possible connection.
We note, however, the number density of the bright LBGs
$n_{\rm LBG}(i'<26.0) \simeq 3.8\times 10^{-3} h^{3}$Mpc$^{-3}$
is in fact comparable to that of OPEGs
$n_{\rm OPEG}(z'<23.8) \simeq 4.7 \times 10^{-3} h^3$Mpc$^{-3}$.
Therefore, we could even argue from this comparison that the
bright LBG--OPEG connection could be the major ancestor--descendant
relation {\it if} the number density of bright LBGs would not
decrease significantly by mergers.

\section*{Acknowledgments}
We would like to thank K. Shimasaku and Y. Suto for valuable
comments and discussions.
This work has been supported in part by a Grant-in-Aid for
Scientific Research (177401166827; 15740126; 14540234) 
of the Ministry of Education, Culture,
Sports, Science and Technology in Japan.
Numerical computations presented in this paper were partly
carried out at ADAC (the Astronomical Data Analysis Center) of the
National Astronomical Observatory, Japan.


\label{lastpage}

\begin{thebibliography}{99}

\bibitem{Adelberger05}
Adelberger K. L., Steidel C. C., Pettini M., Shapley A. E., Reddy N. A., 
Erb D. K., 2005, ApJ, 619, 697

\bibitem{BBKS86}
Bardeen J. M., Bond J. R., Kaiser N., Szalay A. S. 1986,
ApJ, 304, 15

\bibitem{BS01}
Bartelmann M., Schneider P., 2001, Phys. Rep., 340, 291

\bibitem{Bond-etal91}
Bond J. R., Cole S., Efstathiou G., Kaiser N.,
1991, ApJ, 379, 440

\bibitem{Bower91}
Bower R. G., 1991, MNRAS, 248, 332

\bibitem{GALAXEV} Bruzual G., Charlot S., 2003, MNRAS, 344, 1000 

\bibitem{Coil et al 2004}
Coil A. L., Newnman J. A., Kaiser, N., Davis M., Ma C.-P., 
Kocevski D. D., Koo D. C., ApJ, 617, 765

\bibitem{Daddi et al 2001}
Daddi E., Broadhurst T., Zamorani G., Cimatti A., R\"ottgering H.,
Renzini A., 2001, A\&A, 376, 825

\bibitem{Dressler1980}
Dressler A., 1980, ApJ, 236, 351

\bibitem{Firth et al. 2002}
Firth A. E., et al. 2002, 332, 617

\bibitem{Franx03}
Franx M. et al., 2003, ApJ, 587, L79

\bibitem{GD01}
Giavalisco M., Dickinson M., 
2001, ApJ, 550, 177

\bibitem{Groth77} Groth E.~J., Peebles P.~J.~E., 1977, ApJ, 217, 385 

\bibitem{Hamana et al 2004}
Hamana T., Ouchi M., Shimasaku K., Kayo I., Suto Y.,
2004, MNRAS, 347, 813

\bibitem{Kodama et al. 2004}
Kodama T. et al., 2004, MNRAS, 350, 1005

\bibitem{Kravtsov et a; 2004}
Kravtsov A. V., Berlind A., Wechsler R. H., Klypin A. A., Gottl\"ober S., 
Allgood B., Primack J. R., 2004, ApJ, 609, 35

\bibitem{LC93}
Lacey C., Cole S. 1993, MNRAS, 262, 627

\bibitem{ls93}
Landy S. D., Szalay A. S., 1993, ApJ, 412, 64

\bibitem{Maggwick et al 2003}
Maggwick et al., 2003, MNRAS, 344, 847

\bibitem{McCarthy et al 2001}
McCarthy P. J., et al. 2001, ApJL, L131

\bibitem{Miyazaki et al 2003}
Miyazaki M., et al, 2003, PASJ, 55, 1079

\bibitem{MW96}
Mo, H. J., White, S. D. M. 1996, MNRAS, 282, 347

\bibitem{MS02}
Moustakas L.~A., Somerville R.~S., 
2002, ApJ, 577, 1

\bibitem{SXDS-BRiLBG-Ouchietal05}
Ouchi M, et al., 2005, submitted to ApJ, astro, arXiv:astro-ph/0508083 

\bibitem{Ouchi-etal04a}
Ouchi M., et al., 2004a, ApJ, 611, 660

\bibitem{Ouchi-etal04b}
Ouchi M., et al., 2004b, ApJ, 611, 685

\bibitem{PD96} 
Peacock J.~A., Dodds S.~J., 1996, MNRAS, 280, L19

\bibitem{Peebles80}
Peebles P.~J.~E., 1980, The Large-Scale Structure of the Universe 
Princeton Univ. Press, Princeton, NJ

\bibitem{Postman&Geller1984}
Postman M., Geller M. J., 1984, ApJ, 281, 95

\bibitem{PS74} 
Press W.~H., Schechter P., 1974, ApJ, 187, 425

\bibitem{Reddy et al 2005} 
Reddy N. A., Erb D. K., Steidel C. C., Shapley A. E., Adelberger K. L., Pettini M., 
2005, ApJ, in press (astro-ph/0507264)

\bibitem{Roche et al. 2002}
Roche N. R., Almaini O., Dynlop J., Ivison R. J., Willott C. J., 
2002, MNRAS, 337, 1282

\bibitem{Shapley01}
Shapley  A.~E., Steidel C.~C., Adelberger K.~L., 
Dickinson M., Giavalisco M., Pettini M., 2001, ApJ, 562, 95 

\bibitem{ST99}
Sheth R. K., Tormen G., 1999, MNRAS, 308, 119

\bibitem{SXDS-OPEG-Yamadaetal-05}
Yamada T., et al, 2005, ApJ, in press

\bibitem{zehave et al 2002}
Zehavi I., et al, 2002, ApJ, 571, 172

\end{thebibliography}
\end{document}